\documentclass[aps,10pt,prl,twocolumn,showpacs,amssymb,floatfix,superscriptaddress]{revtex4-1}
\usepackage{amsmath,graphics,epsfig}
\usepackage{mathrsfs}
\usepackage{bm}

\newcommand{\mac}{\mathcal}

\newcommand{\tx}{\text}
\newcommand{\ti}{\textit}

\newcommand{\ssf}[1]{\tx{\tiny{#1}}}

\newcommand{\la}{\langle}

\newcommand{\nn}{\nonumber}

\newcommand{\para}{\parallel}
\newcommand{\pr}{\prime}
\newcommand{\ra}{\rangle}
\newcommand{\raw}{\rightarrow}

\newcommand{\alp}{\alpha}
\newcommand{\dlt}{\delta}

\newcommand{\eps}{\epsilon}

\newcommand{\sg}{\sigma}

\newcommand{\ze}{\zeta}

\newcommand{\bn}{\bm{\nabla}} 
\newcommand{\BP}{\hat{\bm{\sigma}}} 
\newcommand{\pmz}{\hat{\sigma}_{z}} 
\newcommand{\bk}{\bm{k}} 
\newcommand{\bkp}{\bm{k}^{\pr}} 
\newcommand{\bo}{\bm{\rho}} 
\newcommand{\np}{n^{\prime}} 

\begin{document}
\title{Scaling Law of Confined Spin-Hall Effect}
\author{Xuhui Wang}
\email[Corresponding author:~]{xuhui.wang@kaust.edu.sa}
\affiliation{King Abdullah University of Science and Technology
(KAUST), Physical Science and Engineering Division, Thuwal
23955-6900, Saudi Arabia}
\author{Jiang Xiao}
\email[Corresponding author:~]{xiaojiang@fudan.edu.cn}
\affiliation{Department of Physics and State Key Laboratory of
Surface Physics, Fudan University, Shanghai 200433, China}
\author{Aur\'{e}lien Manchon}
\affiliation{King Abdullah University of Science and Technology
(KAUST), Physical Science and Engineering Division, Thuwal
23955-6900, Saudi Arabia}
\author{Sadamichi Maekawa}
\affiliation{Advanced Science Research Center, Japan Atomic Energy
Agency, Tokai 319-1195, Japan} \affiliation{CREST, Japan Science
and Technology Agency, Tokyo 102-0075, Japan}
\date{\today}

\begin{abstract}
We incorporate quantum size effect to investigate the extrinsic
spin-Hall effect in ultrathin metal films. A Lippmann-Schwinger
formalism based theoretical method, accounting for quantum
confinement and surface roughness scattering, is developed to
calculate both spin-Hall and longitudinal resistivities and
spin-Hall angle. The presence of quantum confinement gives rise to
a linear relation $\rho_{\ssf{sH}}=\alp \rho+\beta$ between the
extrinsic spin-Hall resistivity $\rho_{\ssf{sH}}$ and longitudinal
charge resistivity $\rho$. The linear term $\alp\rho$ originates
from side jump, and the constant $\beta$ is due to skew
scattering. This deviates significantly from the commonly accepted
scaling law $\rho_{\ssf{sH}}=a\rho^2+b\rho$ in a bulk conductor.
Thus we call for cautious interpretation of experimental data when
applying the scaling law.
\end{abstract}
 \maketitle

Electric manipulation of spin degree of freedom is desirable in
the rising field of spintronics. And spin-Hall effect, generating
a spin current moving transverse to the charge flow without
magnetism or magnetic field, promises a venue offering exactly
that \cite{dp-jetp-1971,she-natmater-2012,she-ieee-2013}. In
nonmagnetic metals, spin-orbit coupled bulk impurity scatters
charge carriers in a spin-selective manner and enables the
spin-Hall effect. This is the so-called extrinsic spin-Hall effect
\cite{hirsch-she,zhang-she} that differs from its intrinsic
sibling which is contingent on particular band structures created
by spin-orbit interaction \cite{tanaka-intrinsic-she-prb-2008,
morota-intrinsic-she-prb-2011}. After decades of theoretical and
experimental assaults, the origins of spin-Hall effect and
anomalous Hall effect (in magnetic metals) are shown to share many
in common. Just as in anomalous Hall effect, theories suggest that
\cite{crepieux-bruno-prb-2001,taka-mae-review,nagaosa-rmp-2010}
the extrinsic spin-Hall effect, too, is dominated by two distinct
mechanisms, skew scattering \cite{smit-skew-scattering-1958} and
side jump \cite{berger-side-jump-1970}. In layman's terms, side
jump describes the spin-dependent deflection of the electron
velocity to opposite directions transverse to the current upon
scattering by an impurity; however, skew scattering is a
manifestation of asymmetric scattering by the spin-orbit coupling
carried by impurities \cite{nagaosa-rmp-2010}.

In most experiments
\cite{kimura-prl-2007,vila-prl-2007,seki-natmater-2008,mosendz-prl-2010,niimi-prl-2011,niimi-prl-2012},
transverse spin-Hall resistivity $\rho_{\ssf{sH}}$ (or
conductivity $\sg_{\ssf{sH}}$) and longitudinal charge resistivity
$\rho$ are often acquired to characterize charge-to-spin
conversion through spin-Hall angle
$\theta_{\ssf{sH}}=\rho_{\ssf{sH}}/\rho$. Apart from the intrinsic
effect not considered here, it is widely accepted that skew
scattering contributes a $\rho_{\ssf{sH}}\propto \rho$, while side
jump offers $\rho_{\ssf{sH}}\propto \rho^{2}$; and the sum yields
an overall scaling law \cite{taka-mae-review}
\begin{align}
\rho_{\ssf{sH}}\approx a \rho^{2}+b\rho.
\label{eq:common-scaling-law}
\end{align}
That such a compact scaling relation bridges the macroscopic
quantities measured in experiments and microscopic processes
proposed by quantum theory makes it one of the central topics in
condensed matter research. In anomalous Hall effect, for example,
this scaling relation is used frequently to parse the underlying
mechanisms from the plots of transverse charge resistivity
$\rho_{\ssf{AH}}$ versus $\rho$ \cite{nagaosa-rmp-2010}.
Temperature and impurity concentration are the common variables
being tuned to vary resistivities and thus acquire the
$\rho_{\ssf{AH}}$ versus $\rho$ curves. Yet another elegant
experimental paradigm, as proposed recently by Tian \ti{et al.}
\cite{tian-prl-2009}, opted for the sample thickness as the
adjustable parameter to achieve changes in resistivities. Their
results, however, call for more meticulous review on the
application of the existing scaling law in an anomalous Hall
system.

Indeed, theoretical effort that concludes with relation
(\ref{eq:common-scaling-law}) often assumes a bulk conductor
\cite{nagaosa-rmp-2010,taka-mae-review}. But modern-day spin-Hall
or anomalous Hall experiments are usually performed in ultrathin
films as thin as a few monolayers. This juxtaposition no longer
justifies the negligence of confinement and surface roughness.
Early transport experiments and theories have already pointed out
an important phenomenon that charge conductivity in thin films can
be modified dramatically by quantum confinement coupled with
surface roughness scattering
\cite{hensel-prl-1985,tjm-prl-1986,trivedi-prb-1988}. More recent
studies in combining confinement with spin active surfaces propose
unconventional ways to generate spin-Hall effect by either
interfacial spin-orbit coupling
\cite{wang-prbrc-2013,borge-prb-2014} or even surface roughness
\cite{zhou-unpub-2014}. We are therefore much intrigued by
possible novelties arising from merging the quantum size effect
and spin-Hall phenomenon.

In this Letter, we incorporate quantum confinement and surface
roughness into a theoretical analysis of the extrinsic spin-Hall
effect. We discover that the spin-Hall resistivities
(conductivities) due to side jump and skew scattering, in the
presence of quantum confinement, acquire different thickness
dependencies. More importantly, for the change in resistivities
measured in experiments as a result of varying thickness, the
quantum confinement provides a linear scaling law
$\rho_{\ssf{sH}}\approx \alp \rho + \beta$, deviating
significantly from formula (\ref{eq:common-scaling-law}). We must
therefore emphasize that parsing the underlying physical
mechanisms using the existing scaling law shall be carried out
with caution. Moreover, we find that the spin-Hall angle can be
tuned by surface roughness fluctuation. This may provide an
alternative route, in addition to tuning impurity concentration
\cite{niimi-prl-2012}, towards an efficient control of the
charge-spin interconversion.

We are primarily interested in nonmagnetic metals that accommodate
spin-orbit coupled impurity and thus robust extrinsic spin-Hall
effect \cite{kimura-prl-2007,seki-natmater-2008}. Our model
includes a bulk spin-orbit scattering potential $V_{\ssf{SO}} =
-i\eta_{\ssf{SO}}\BP\cdot\left(\bn V_{\ssf{I}}\times\bn\right)$,
generated as a relativistic correction to $\dlt$-type nonmagnetic
impurities
$V_{\ssf{I}}=V_{\ssf{imp}}\sum_{i}\dlt(\bm{r}-\bm{r}_{i})$ located
at $\bm{r}_{i}$ \cite{taka-mae-review}. We assume that
$V_{\ssf{I}}$ has a magnitude $V_{\ssf{imp}}$ and impurity
concentration $n_{i}$. $\eta_{\ssf{SO}}$ is the spin-orbit
coupling constant and $\BP$ the Pauli matrix. In the free electron
model, the impurity scattering potential $V_{\ssf{I}}$ gives rise
to the well-known transport relaxation rate $\tau_{\ssf{0}}^{-1}=m
V_{\ssf{imp}}^{2}n_{i}k_{F}/\pi\hbar^{3}$ and Drude conductivity
$\sg_{\ssf{0}}=e^{2}\tau_{\ssf{0}}n_{e}/m$, where $n_{e}$ is the
electron density \cite{ziman-book}.

To introduce quantum confinement, we consider the metal film to
have an average thickness $d$ and is terminated by two infinite
potential barriers at two surfaces. From the energy point of view,
the film is then modelled as a square well potential $U_{0}(z)$
\cite{tjm-prl-1986, trivedi-prb-1988}. For an electron with
effective mass $m$, its motion along the confinement direction
$\hat{\bm{z}}$, described by a Hamiltonian
$H_{0,\perp}=\hbar^{2}k_{z}^{2}/2m$, is quantized into
$n_{c}=\lfloor k_{F}d/\pi\rfloor$ conducting channels. Here,
$k_{F}$ is the Fermi wave vector and the floor function $\lfloor a
\rfloor$ gives the largest integer not greater than $a$. On the
other hand, the motion in the $xy$ plane is captured by a
Hamiltonian $H_{0,\para}=\hbar^{2}(k_{x}^{2}+k_{y}^{2})/2m$. A
free electron state of spin-$s$ in channel $n$ is thus represented
by a wave function
\begin{align}
|\bk n s\ra
=\sqrt{\frac{2}{\mac{V}}}\sin\left(\frac{n\pi}{d}z\right)
e^{i\bo\cdot\bk}|s\ra \label{eq:free-state}
\end{align}
together with an energy eigenvalue $E_{\bk n s}$ that fulfills
$(H_{0,\para}+H_{0,\perp}+U_{0})|\bk n s\ra=E_{\bk n s}|\bk n
s\ra$. In wave function (\ref{eq:free-state}), $\bo$ and
$\bk=(k_{x},k_{y})$ are the in-plane coordinate and momentum,
respectively, whereas $\mac{V}$ is the volume of the film. It is
worth to point out that, due to the confinement, the density of
states at Fermi energy becomes $N_{F}=m n_{c}/2\pi\hbar^{2}d$, and
is therefore only weakly dependent on thickness through $n_{c}$.

At this stage, we may ask two seemingly simple questions: (i) How
does the quantum size effect impact the side-jump and
skew-scattering processes? (ii) Do we expect any change in the
scaling relation (\ref{eq:common-scaling-law})? One might try to
argue that, as $N_{F}$ shows, the discretization in conducting
channels introduces size dependence in the density of states,
which is translated into thickness dependence in the scattering
rate and thus the resistivity subscribing to either side jump or
skew scattering. This appears to influence only the magnitude of
$\rho_{\ssf{sH}}$ and $\rho$, but not the formal structure of
relation (\ref{eq:common-scaling-law}). Indeed, the transport
relaxation rate due to bulk impurity alone in the presence of
quantum confinement is \cite{trivedi-prb-1988}
\begin{align}
\frac{1}{\tau_{\ssf{I}}} =\frac{m
V_{\ssf{imp}}^{2}n_{i}}{\hbar^{3}}
\frac{\left(n_{c}+\frac{1}{2}\right)}{d},
\label{eq:scattering-rate-side-jump}
\end{align}
where a small contribution due to spin-orbit impurities,
proportional to $\eta_{\ssf{SO}}^{2}$, has been neglected. In
comparison with the rate $\tau_{0}^{-1}$ in the bulk,
$\tau_{\ssf{I}}^{-1}$ exhibits a weak size effect through the
density of states. This is not the entire story, though. As to be
disclosed in the rest, a na\"{i}ve picture of such, however, does
not survive serious scrutiny.

First, in an ultrathin film, we must take the role of the surface
roughness into account. An uneven surface along the transport
direction means a continuous compression or dilation of the
transverse wave function, causing exchange between the in-plane
and transverse energies, which results in an additional effective
scattering process. Such effect is irrelevant when we consider a
bulk conductor. But in miniature structures delivered by
state-of-the-art nanotechnology, this surface roughness scattering
reduces the longitudinal charge conductivity and generates
non-trivial thickness dependence \cite{tjm-prl-1986,
trivedi-prb-1988}. To treat a small-scale surface roughness
theoretically, we apply a nonunitary dilation operator
$e^{\ze(\bo)}e^{i\ze(\bo)(zp_{z}+p_{z}z)/(2\hbar)}$ to convert the
film of a constant surface into one with a random surface
\cite{tjm-prl-1986, trivedi-prb-1988}. The surface roughness
considered here has a white-noise profile with a standard
deviation $\dlt_{\tx{t}}$ \cite{surface-roughness}. As a result of
the dilation operation and to the leading order in thickness
fluctuation $\ze$ and $\eta_{\ssf{SO}}$, the total Hamiltonian for
the think film becomes
\begin{align}
H_{\ssf{R}}=H_{0,\para}+H_{0,\perp}(z)+U_{0}(z)+V_{\ssf{I}}+V_{\ssf{SO}}+V_{\ssf{R}},
\end{align}
i.e., augmented by an additional spin-independent pseudo potential
$V_{\ssf{R}}=2\ze H_{0,\perp}+[S_{0},H_{0,\perp}]$, where
$S_{0}=i\ze(\bo)(zp_{z}+p_{z}z)/(2\hbar)$, to be treated
perturbatively. Here, we must emphasize that neither the bulk
impurity potential $V_{\ssf{I}}$ nor the spin-orbit coupling
$V_{\ssf{SO}}$ is invariant under the dilation transformation, but
the effects are of higher order in both $\ze$ and
$\eta_{\ssf{SO}}$ and are thus discarded.

In order to pursue spin-Hall conductivity in the presence of both
quantum confinement and surface roughness, we must establish a
theoretical tool that is able to treat side jump as well as skew
scattering in one setting. We introduce an operator
$\hat{j}_{x}^{z}= ({e}/{4})\left\{\hat{v}_{x},\pmz\right\}$ for
the spin current flowing along the $\hat{\bm{x}}$ direction and
polarized along $\hat{\bm{z}}$ \cite{sinova-prl-she-2004}. The
velocity operator $\hat{v}_{x} = p_{x}/m +
(\eta_{\ssf{SO}}/\hbar)(\BP\times\bn V_{\ssf{I}})_{x}$, having
both normal and anomalous components, is derived from Heisenberg
equation. With the free electron wave function
(\ref{eq:free-state}), we construct, using Lippmann-Schwinger
equation \cite{taka-mae-review}, a scattered state
\begin{align}
|\phi_{\bk n s}^{+}\ra =|\bk n s\ra+\sum_{\bkp,n^{\pr}}|\bkp
n^{\pr} s\ra \frac{\la\bkp \np s|V_{\ssf{I}}+V_{\ssf{R}}|\bk
ns\ra}{E_{\bk n}-E_{\bkp n^{\pr}}+i\eps},
\label{eq:scattering-state-confinement}
\end{align}
where symbol $\eps$ is a positive infinitesimal. Transition
probability rate is therefore provided by Fermi's golden rule
\begin{align}
P_{\bkp \np s^{\pr};\bk n s} & \nn\\ =\frac{2\pi}{\hbar} &
\la\la|\la \bkp \np s^{\pr}|\hat{T}|\bk n s\ra|^{2}\ra\ra
\dlt(E_{\bkp \np s^{\pr}}-E_{\bk n s}) \label{eq:fermi-golde-rule}
\end{align}
with the $T$-matrix $\hat{T} =
\hat{U}+\hat{U}(E-\hat{H}_{0})^{-1}\hat{T}$ that invokes the
\ti{full} scattering potential
$\hat{U}=V_{\ssf{I}}+V_{\ssf{SO}}+V_{\ssf{R}}$. Symbol
$\la\la\cdots\ra\ra$ denotes the configurational average over both
bulk impurity and surface roughness profile. In principle,
transport and spin relaxation rates can be obtained from
probability (\ref{eq:fermi-golde-rule}).

When an electric field $\bm{\mac{E}}$ is applied along the
transport direction $\hat{\bm{y}}$, the spin-Hall conductivity is
defined as $\sg_{\ssf{sH}}= -J_{x}^{z}/\mac{E}$ and the
ensemble-averaged spin current is obtained from
\begin{align}
J_{x}^{z}=\frac{1}{\mac{V}}\sum_{\bk n s}f_{\bk n s}
\la\la\la\phi_{\bk n s}^{+}|\hat{j}_{x}^{z}|\phi_{\bk n
s}^{+}\ra\ra\ra,
\end{align}
where $f_{\bk n s}$ is the distribution function. It is
sufficient, in the present calculation, to partition the
distribution function into $f_{\bk n s}=f_{\bk n s}^{(0)}+f_{\bk n
s}^{(1)}+f_{\bk n s}^{(2)}$, where $f_{\bk n s}^{(0)}$ is the
equilibrium Fermi-Dirac distribution, $f_{\bk n s}^{(1)}$ is the
first-order correction due to momentum relaxation, and $f_{\bk n
s}^{(2)}$ is the second-order correction attributed to skew
scattering. The framework outlined above is different from
previous approaches using either Kubo formalism
\cite{tjm-prl-1986} or density matrix \cite{trivedi-prb-1988}, and
in line with the method developed by Takahashi and Maekawa
\cite{taka-mae-review}.

The localization effect and interference between scattering events
are neglected, which is justified by a low impurity concentration
and weak scattering at impurities or surface roughness. Surface
roughness, as an independent scattering mechanism, contributes a
channel-dependent relaxation rate \cite{tjm-prl-1986,
trivedi-prb-1988}
\begin{align}
\frac{1}{\tau_{\ssf{R},n}}=\frac{2 E_{F}}{\hbar }
\left(\frac{\dlt_{\tx{t}}}{d}\right)^{2}
\frac{n^{2}}{n_{c}^{4}}\sum_{\np=1}^{n_{c}}{\np}^{2}.
\label{eq:scattering-rate-roughness}
\end{align}
We see that $\tau_{\ssf{R},n}$ is proportional to $d^{3}$, i.e.,
having a much more dramatic change with the thickness than
$\tau_{\ssf{I}}$. The total relaxation rate in the thin film is
given by
$\tau_{n}^{-1}=\tau_{\ssf{I}}^{-1}+\tau_{\ssf{R},n}^{-1}$. The
longitudinal (charge) conductivity is \cite{trivedi-prb-1988}
\begin{align}
\sg = \sg_{\ssf{0}} \frac{3}{2
n_{c}}\sum_{n}\frac{\tau_{n}}{\tau_{\ssf{0}}}
\left(1-\frac{E_{0}}{E_{F}}n^{2}\right),
\label{eq:longitudinal-conductivity}
\end{align}
where $E_{0}=\hbar^{2}/2md^{2}$, and reduces to the Drude
conductivity $\sg_{\ssf{0}}$ when both confinement and roughness
are removed \cite{ziman-book}.

We divide the spin-Hall conductivity into the one that is due to
side jump and the other coming from skew scattering
\cite{taka-mae-review}. For the side-jump contribution, we only
need to consider the first-order correction to the distribution
function $f_{\bk n s}^{(1)}\approx (e\hbar/m)\tau_{n}
\bm{\mac{E}}\cdot\bm{k}\dlt(E_{\bk n}-E_{F})$. This gives a
spin-Hall conductivity
\begin{align}
\sg_{\ssf{SJ}} = \alp_{\ssf{SJ}}\sg, \label{eq:cond-side-jump}
\end{align}
where $\alp_{\ssf{SJ}}={\eta_{\ssf{SO}} m}/{\hbar \tau_{\ssf{I}}}$
is the dimensionless side-jump parameter \cite{lyo-holstein,
taka-mae-review}. Equation (\ref{eq:cond-side-jump}) shows that
the side-jump induced spin-Hall conductivity shall have the same
thickness dependence as $\sg$. This means, when the surface
roughness scattering dominates the relaxation, or equivalently
$\tau_{\ssf{R},n}^{-1}\gg \tau_{\ssf{I}}^{-1}$, the spin-Hall
conductivity $\sg_{\ssf{SJ}}\sim (d/\dlt_{\tx{t}})^{2}$.

The skew scattering in spin-Hall effect requires the second-order
correction to the distribution function and thus the knowledge of
the antisymmetric transition probability \cite{taka-mae-review}.
However, we find that surface roughness does not alter the
antisymmetric transition probability which is merely due to the
bulk impurity $V_{\ssf{I}}$ and $V_{\ssf{SO}}$
\begin{align}
&P_{\bkp n^{\pr}s^{\pr};\bk n s}^{\ssf{(a)}}\nn\\
&=\frac{2\pi s}{\hbar}\dlt_{s s^{\pr}}\eta_{\ssf{SO}} \frac{m
V_{\ssf{imp}}^{3}n_{i}n_{c}}{2 \hbar^{2}\mac{V}d}
(\bk\times\bkp)_{z} \dlt(E_{\bk n}-E_{\bkp n^{\pr}}).
\label{eq:asymmetric-transition-rate}
\end{align}
Interestingly, even in the opposite scenario where the bulk
spin-orbit impurities vanish yet the surface roughness provides
the only spin-orbit coupling, the surface roughness still makes no
contribution to the antisymmetric transition probability
\cite{zhou-unpub-2014}. We further insert the transition rate
(\ref{eq:asymmetric-transition-rate}) into the Boltzmann equation
to obtain
\begin{align}
f_{\bk n s}^{(2)}=-\frac{e\hbar}{m} \tau_{n} \sum_{\bkp
n^{\pr}s^{\pr}} & P_{\bkp n^{\pr} s^{\pr};\bk n
s}^{\ssf{(a)}}\nn\\
&\times \tau_{n^{\pr}} \bm{\mac{E}}\cdot\bm{k}^{\pr}\dlt(E_{\bkp
n^{\pr}}-E_{F}).
\end{align}
It leads to the desired spin-Hall conductivity due to skew
scattering, in the presence of confinement and surface roughness,
\begin{align}
\sg_{\ssf{SS}} = \frac{\beta_{\ssf{SS}}}{\sg_{0}}\sg^{2}
\end{align}
where
$\beta_{\ssf{SS}}=({\pi}/{2\hbar^{2}})m\eta_{\ssf{SO}}n_{e}V_{\ssf{imp}}$.
If we compare the last result to conductivities
(\ref{eq:longitudinal-conductivity}) and
(\ref{eq:cond-side-jump}), we see that the spin-Hall conductivity
produced by skew scattering has a much more prominent thickness
dependence $\sg_{\ssf{SS}}\sim (d/\dlt_{\tx{t}})^{4}$, so long as
the roughness scattering dominates. The total spin-Hall
conductivity shall combine the contributions from two mechanisms,
i.e., $\sg_{\ssf{sH}}=\sg_{\ssf{SJ}}+\sg_{\ssf{SS}}$. That side
jump and skew scattering depend on thickness and surface roughness
in different manners actually points to an alternative method to
distinguish the underlying physical mechanisms driving the
extrinsic spin-Hall effect.

On the other hand, the spin-Hall resistivity is a quantity
measured frequently. In terms of longitudinal resistivity
$\rho=\sg^{-1}$, it becomes
\begin{align}
\rho_{\ssf{sH}}\approx \frac{\sg_{\ssf{sH}}}{\sg^{2}}
=\alp_{\ssf{SJ}}\rho+\beta_{\ssf{SS}}\rho_{0},
\label{eq:spin-hall-resistivity}
\end{align}
where $\rho_{0}=\sg_{0}^{-1}$ is the bulk Drude resistivity. The
last expression (\ref{eq:spin-hall-resistivity}), as the central
result of this Letter, deserves a thorough discussion since it is
highly relevant to most experiments. First, it is quintessential
to realize that, in any experiment, the resistivities being
actually measured are $\rho_{\ssf{sH}}$ and $\rho$, i.e. the ones
that shall subscribe to both impurity and roughness scatterings
(thus the confinement effect). They shall be distinguished from
the ideal bulk value $\rho_{0}$. This fact has already been
demonstrated by earlier experiments in thin films
\cite{hensel-prl-1985,tjm-prl-1986,trivedi-prb-1988}. Furthermore,
the size dependence of $\rho_{\ssf{sH}}$ is governed entirely by
$\rho$, since $\alp_{\ssf{SJ}}$, $\beta_{\ssf{SS}}$ and $\rho_{0}$
are independent of thickness and surface roughness.

The above discussion leads us to consider one experimental
paradigm which is rather practical in reality
\cite{tian-prl-2009}. In order to tune the resistivity of the thin
film, we \ti{only} change its thickness or the surface roughness
while keeping other parameters--such as material, doping
concentration, and temperature--untouched. In this way,
$\rho_{\ssf{sH}}$ and $\rho$ will change accordingly, while
$\alp_{\ssf{SJ}}$, $\beta_{\ssf{SS}}$ and $\rho_{0}$ shall remain
constant. When plotting the curve of $\rho_{\ssf{sH}}$ versus
$\rho$, we therefore see a \ti{linear} relation as
(\ref{eq:spin-hall-resistivity}) instead of
$\rho_{\ssf{sH}}=a\rho^{2}+b\rho$ which contains a \ti{quadratic}
term. In this $\rho_{\ssf{sH}}$ vs $\rho$ plot, there is a nonzero
intercept (or residual resistivity) on the $\rho_{\ssf{sH}}$ axis
as $\rho\raw 0$, which traces back to the skew scattering. Most
importantly, in scaling law (\ref{eq:spin-hall-resistivity}), the
linear component $\alp_{\ssf{SS}}\rho$ is a result of side-jump
mechanism, \ti{not} due to skew scattering, which is the contrary
to what is usually interpreted using relation
(\ref{eq:common-scaling-law}). As confinement becomes negligible
in a bulk conductor, namely, the quantum size effect and role of
surface roughness diminish, the value of $\rho$ acquired in
experiment may approach the ideal Drude value of $\rho_{\ssf{0}}$.
And it is only in this limit, we are allowed to accept the
interpretation based on scaling law
$\rho_{\ssf{sH}}=a\rho^{2}+b\rho$. It is worth to note that, our
analysis in this Letter is done for a system without magnetism, we
nevertheless emphasize the importance to properly account for the
quantum size effect and surface roughness when magnetism is
present in, for example, anomalous Hall effect.

Another interesting quantity is the spin-Hall angle usually
defined as $\theta_{\ssf{sH}}=\rho_{\ssf{sH}}/\rho$. As a
by-product of the scaling law (\ref{eq:spin-hall-resistivity}), it
becomes $\theta_{\ssf{sH}} =
\alp_{\ssf{SJ}}+\beta_{\ssf{SS}}\rho_{0}\sg$, indicating the
thickness dependence of $\theta_{\ssf{sH}}$ follows that of $\sg$.
This relation suggests an alternative route to achieve desired
$\theta_{\ssf{sH}}$ by altering the ratio of thickness fluctuation
$\dlt_{\tx{t}}$ to thickness $d$, since the present experimental
techniques are likely to offer better control in $\dlt_{\tx{t}}$
and $d$ than in many other parameters such as $n_{i}$ or
$V_{\ssf{imp}}$.

We conclude the paper by asserting that, in spin-Hall systems
constructed on ultrathin films, quantum confinement and surface
roughness scattering induce thickness dependence which shall not
be ignored. Such quantum size effects are embedded in the
resistivities measured experimentally and the scaling law thus
deviates from the one derived for a bulk conductor. Therefore,
more caution shall be exercised in the interpretation of
experimental data using the existing scaling relation,
particularly in the case when the tuning of resistivity is
accomplished by changing the film thickness. The influence of
quantum confinement on the intrinsic contribution to spin-Hall
effect is beyond the scope of this paper. As a final remark, we
speculate that the reduction in phase space due to confinement has
a rather limited impact on the Bloch states and thus the band
degeneracy needed for intrinsic effect. We thus envisage the
intrinsic effect to lead the contribution as $\rho^{2}$ in the
scaling law. To rigourously reveal this for a real-world material,
the existing numerical schemes
\cite{tanaka-intrinsic-she-prb-2008,morota-intrinsic-she-prb-2011,lowitzer-prl-2011}
must be supplemented with a proper implementation of quantum
confinement.

X. Wang and A. Manchon acknowledge the support from the King
Abdullah University of Science and Technology (KAUST). J. Xiao
acknowledges the support from the special funds for the Major
State Basic Research Project of China (2014CB921600, 2011CB925601)
and the National Natural Science Foundation of China (91121002).


\begin{thebibliography}{999}
\bibitem{surface-roughness}
The thickness fluctuation $\ze$ has a vanishing mean and a finite
second moment. Then the roughness profile average on the Fourier
components of $\ze$ is $\la\ze_{\bm{q}}^{\ast}\ze_{\bm{q}}\ra
=(1/\mac{A}k_{F}^{2})(\dlt_{\tx{t}}/d)^{2}$, where $\mac{A}$ is
the area of the surface.
\end{thebibliography}
\end{document}